\documentclass[aip,jmp,twocolumn,amsmath,amssymb,showpacs]{revtex4-1}
\usepackage{amssymb}
\usepackage{epsfig}
\usepackage{amsmath}
\usepackage{times}

\begin{document}

\title{An efficient approach to suppress the negative role of
contrarian oscillators in synchronization}

\author{Xiyun Zhang}
\affiliation{Department of Physics, East China Normal University,
Shanghai, 200062, P. R. China}

\author{Zhongyuan Ruan}
\affiliation{Department of Physics, East China Normal University,
Shanghai, 200062, P. R. China}

\author{Zonghua Liu}
\email{zhliu@phy.ecnu.edu.cn} \affiliation{Department of Physics,
East China Normal University, Shanghai, 200062, P. R. China}

\date{\today}

\begin{abstract}
It has been found that contrarian oscillators usually take a negative
role in the collective behaviors formed by conformist oscillators.
However, experiments revealed that it is also possible to achieve a
strong coherence even when there are contrarians in the system such as
neuron networks with both excitable and inhibitory neurons. To understand
the underlying mechanism of this abnormal phenomenon, we here consider
a complex network of coupled Kuramoto oscillators with mixed positive and
negative couplings and present an efficient approach, i.e. tit-for-tat
strategy, to suppress the negative role of contrarian oscillators in
synchronization and thus increase the order parameter of synchronization.
Two classes of contrarian oscillators are numerically studied and a brief
theoretical analysis is provided to explain the numerical results.

\end{abstract}



\maketitle

\begin{quotation}
Synchronization represents the emergence of macroscopic state of an ensemble of
coupled oscillators and focuses recently on the influence of network topology.
Generally, the considered system consists of identical or non-identical oscillators
with an attractive coupling. However, realistic systems may be more complicated
with both positive and negative couplings. To understand the role of the negative
coupling, some preliminary studies have revealed that the negative couplings
usually suppress the synchronization, which may be important to the control of
undesired synchronization such as the abnormal synchronization of epileptic
seizure. On the other hand, the negative coupling may also take positive role
in sustaining the functions of systems. For example, in neuron networks, the
existence of both excitable and inhibitory neurons is necessary for the formation
of different memories. In animals, the competition between cooperation and defection
results in the co-existence or diversity of animals. In this paper, we try to
understand the microscopic mechanism of the positive role of the negative coupling.
We reveal that a possible way to implement the positive role of the negative
coupling is the tit-for-tat strategy, which may open a new window to study the
positive roles of contrarians.
\end{quotation}

\section{Introduction}
Networks are ubiquitous in both natural and technical systems. The
dynamics of coupled oscillators in complex networks has obtained
an increasing interesting. It is found that the collective
behaviors of coupled oscillators depend on the coupling, network
topology, and the intrinsic properties of individual oscillators \cite{Albert:2002,Dorogovtsev:2008,Arenas:2008}.
Most of the studies are focused on the case of the same
kind of oscillators. However, considering that real systems are
sometimes very complicated, the studies on mixed oscillators with
two kinds of couplings, i.e. conformists and contrarians, has
recently gotten a great attention
\cite{Strogatz:2011,Hong:2011,Louzada:2012,Restrepo:2005,Zanette:2005,Borgers:2003}.
The conformists are represented by a positive coupling strength while
contrarians by a negative coupling strength. It is believed that the
positive coupling will force the oscillators to become synchronized while
the negative coupling will drive the oscillators to be out of phase.

A simple model to study the collective behaviors of coupled
oscillators is the Kuramoto model. When mixed positive and
negative couplings are considered, the Kuramoto model can show a
glass transition and can be also used to describe the neural
networks \cite{Daido:1992,Bonilla:1993,Zanette:2005}. For the
former, the positive coupling denotes a ferromagnetic interaction
while the negative coupling denotes an anti-ferromagnetic
interaction. When both types of couplings are present, the system
becomes frustrated. For the latter, the positive coupling
represents an excitatory coupling while the negative coupling
represents an inhibitory coupling. In most real systems, there are
contrarians although they are the minority. For example, in
political election or rumor propagation, we have to face the
contrarians. In brain, it is well known that the fraction of
excitable neurons is about $75\%$ while that of inhibitory neurons
is about $25\%$ \cite{Soriano:2008,Vogels:2009}. Several models
have been presented to understand the role of contrarians. It is
revealed that the contrarians can lead to rich dynamics such as
two diametrically opposed factions, traveling waves, and complete
incoherence etc \cite{Strogatz:2011,Hong:2011}. It is also
revealed that local contrarians can be used to suppress undesired
synchronization \cite{Louzada:2012}. In sum, the contrarians
usually take a negative role in the formation of collective
behaviors.

On the other hand, we see the evidences that a highly coherence
can be finally formed although there exist some contrarians in the
system. For example, a final winner will be generated in political
election with contrarians. In neural networks with both excitable
and inhibitory neurons, experiments have shown the large-scale
cortical synchronization \cite{Hipp:2011,Roelfsema:1997} and
synfire propagation in cognitive process where the signal is
carried by a wave of synchronous neuronal activity within a subset
of network neurons
\cite{Diesmann:1999,Singer:1999,Vogels:2005,Womelsdorf:2007}.
These results imply that realistic systems may have their ways to
suppress the negative role of contrarians. Then, an interesting
question is how they do that or what is the possible way to do
that. To answer these questions, we here try to present an
approach to implement the suppression of contrarians in a
mathematical model such as in the Kuramoto model with mixed
positive and negative couplings.

In this paper, we consider a complex network of coupled Kuramoto
oscillators with mixed positive and negative couplings and present
an efficient approach, i.e. tit-for-tat strategy, to suppress the
negative role of contrarian oscillators in synchronization. We focus
on two classes of contrarian oscillators and numerically find that the
tit-for-tat strategy can increase the order parameter of synchronization for
both classes. A brief theoretical analysis is provided to explain the
numerical results. In fact, this approach of tit-for-tat has been widely
used in the aspect of aggression in animals \cite{Akcay:2009},
humans \cite{Kramer:2007}, as well as in the social interaction among
different bacterial species \cite{Basler:2013}, to enforce collaboration
among selfish users. But a theoretical model to explain its microscopic
mechanism is still lacking so far. The tit-for-tat strategy has two rules:
(1) cooperate if your partner cooperates and (2) defect if your partner
defects \cite{Axelrod:1981,Godard:1993,Milinski:1987}. We here use
these two rules to the Kuramoto model with mixed positive and negative
couplings.

In the previous studies, a contrarian oscillator can be
distinguished from a conformist oscillator by the coupling term in
two ways. In the first way, a contrarian oscillator will receive
interactions from its neighbors by a negative coupling strength
while a conformist oscillator will receive interactions from its
neighbors by a positive coupling strength
\cite{Strogatz:2011,Hong:2011,Louzada:2012}. In the second way,
a contrarian oscillator will give negative coupling to each of its
neighbors while a conformist oscillator will give a positive coupling
to each of its neighbors \cite{Restrepo:2005,Zanette:2005,Borgers:2003}.
We here first discuss these two cases, respectively, and then
theoretically show the mechanism of tit-for-tat strategy in coupled
Kuramoto oscillators.

The paper is organized as follows. In Sec. II, we present the
model and discuss the {\em Case I} of contrarian oscillators
receiving couplings with negative coupling strength. Then in Sec.
III, we discuss the {\em Case II} of contrarian oscillators
sending negative coupling to their neighbors. After that, we
present a brief theoretical analysis to explain the numerical
results in Sec. IV. Finally we summarize our results in Sec. V.

\section{Case I: contrarian oscillators receiving couplings with
negative coupling strength}

We consider a network of $N$ coupled Kuramoto oscillators. Each
oscillator is characterized by its phase $\theta_i(t), i =
1,\cdots,N$ and obeys an equation of motion defined as
\begin{equation}
\label{Kuramoto}
\dot{\theta_{i}}=\omega_i+\lambda_i\sum_{j=1}^NA_{ij}\sin(\theta_j-\theta_i),
\quad i=1,\ldots,N
\end{equation}
where $\lambda_i$ is the coupling strength of the oscillator $i$,
$\omega_i$ is its natural frequency, and
$A_{ij}$ are the elements of the adjacency matrix $A$, so that
$A_{ij}=1$ when nodes $i$ and $j$ are connected and $A_{ij}=0$
otherwise. For simplicity, we let all the coupling strength
$\lambda_i$ have the same amplitude $\lambda>0$ with
$\lambda_i=\lambda$ for all the conformists and $\lambda_i=-\lambda$
for all the contrarians. In the network, the contrarians will be
random uniformly mixed with the conformists. We use $\rho$ to
represent the ratio between the number of the contrarians and the
total number of oscillators, i.e. $\rho N$ contrarians and $(1-\rho) N$
conformists in the network.

To measure the coherence of the collective motion, we use
the order parameter $R$ \cite{Restrepo:2005,Jesus:2011}:
\begin{equation}
\label{order} Re^{i\Psi}=\frac{1}{N}\sum_{j=1}^N e^{i\theta_j},
\end{equation}
where $\Psi$ denotes the average phase, and $R$ ($0\leq R \leq 1$)
is a measure of phase coherence. $R$ will reach unity when the
system is fully synchronized and be $0$ for an incoherent
solution. As the conformists will attract each other to form a
synchronized cluster and the contrarians will make each other be
out of phase, the final formed collective behavior will depend on
the competition between the conformists and the contrarians. In
this sense, the contrarians will take a negative role to the final
formed collective behavior. There are three kinds of links in the
network: (1) conformist-conformist; (2) contrarian-contrarian; and
(3) conformist-contrarian. The first two kinds of links are
symmetric while the third one is asymmetric. To increase the order
parameter $R$ of synchronization, our idea is to let the
asymmetric links of conformist-contrarian (the third one) become a
symmetric link of contrarian-contrarian (the second one) by a
probability $P$, i.e. the tit-for-tat strategy. This strategy
originates from the game theory such as the Prisoner's Dilemma where
two individuals can each either cooperate or defect
\cite{Axelrod:1981,Godard:1993,Milinski:1987}. In this theory, the
payoff to a player is in terms of the effect on its fitness. No matter
what the other does, the selfish choice of defection yields a higher
payoff than cooperation. But if both defect, both do worse than if both
had cooperated. Thus, a safe strategy is to follow the simple rule of
copying the last behavior of its partner. In this sense, the tit-for-tat
strategy is proposed to enforce collaboration. In detail, we
first find all the asymmetric links with $\lambda_iA_{ij}
\not=\lambda_jA_{ji}$. Then we let the positive one in the pair of
$\lambda_iA_{ij}$ and $\lambda_jA_{ij}$ become negative by the
probability $P$.

\begin{figure}
\epsfig{figure=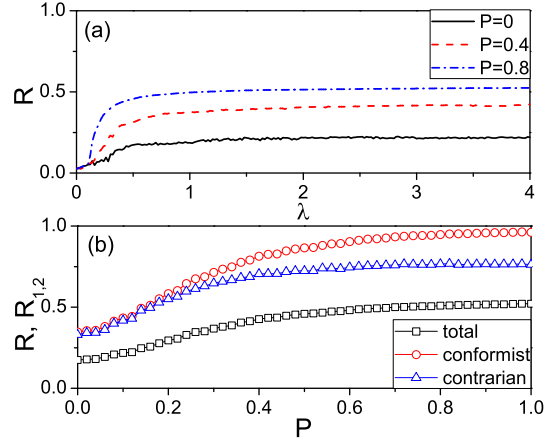,width=1.0\linewidth} \caption{(color
online). Synchronization diagrams by the tit-for-tat strategy. (a)
$R$ versus $\lambda$ where the ``solid", ``dashed", and
``dash-dotted" lines represent the cases of $P=0, 0.4$, and $0.8$,
respectively. (b) $R, R_1$ and $R_{2}$ versus $P$ for $\lambda=3.0$,
where the lines with ``squares", ``circles", and ``triangles"
represent the cases of the total oscillators, the group of
conformists, and the group of contrarians, respectively.}
\label{Fig:order-parameter}
\end{figure}

In numerical simulations, we first take the random Erdos-Renyi
(ER) network with the average degree $\langle k\rangle=6$ as an
example. We let the frequency $\omega_i$ in Eq. (\ref{Kuramoto})
satisfy the Lorentzian distribution of
$g(\omega)=\frac{1}{\pi}[\frac{\gamma}{(\omega-\omega_0)^2+\gamma^2}]$
with the central frequency $\omega_0$ and $\gamma$ is the half
width at half maximum \cite{Matthews:1991}. We let the parameters
be $N=1000$, $\omega_0=0$, $\gamma=0.2$ and $\rho=0.25$ in this
paper. Fig. \ref{Fig:order-parameter}(a) shows how $R$ changes
with $\lambda$ where the ``solid", ``dashed", and ``dash-dotted"
lines represent the cases of $P=0, 0.4$, and $0.8$, respectively.
It is easy to see that the three lines are overlapped when
$\lambda$ is very small and then they quickly separated when
$\lambda$ is gradually increased. The ``squares" in Fig.
\ref{Fig:order-parameter}(b) shows how $R$ depends on $P$. We see
that $R$ gradually increases with $P$, indicating that the
tit-for-tat strategy can successfully suppress the negative role
of contrarians in synchronization.

\begin{figure}
\epsfig{figure=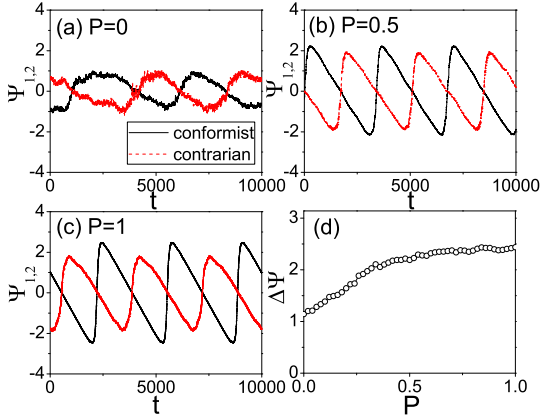,width=1.0\linewidth} \caption{(color
online). Evolution of the average phases $\Psi_1$ and $\Psi_2$
where (a)-(c) represent the cases of $P=0, 0.5$ and $1$,
respectively, and (d) represents the dependence of $\Delta \Psi$
on $P$.} \label{Fig:phase}
\end{figure}

To reveal the underlying mechanism of tit-for-tat strategy, we
introduce two new order parameters $R_1$ and $R_2$ defined by
\begin{equation}
\label{order-conformist}
R_1e^{i\Psi_1}=\frac{1}{(1-\rho)N}\sum_{j=1}^{(1-\rho)N}
e^{i\theta_j}
\end{equation}
for the group of conformists and
\begin{equation}
\label{order-contrarian} R_2e^{i\Psi_2}=\frac{1}{\rho
N}\sum_{j=1}^{\rho N} e^{i\theta_j}
\end{equation}
for the group of contrarians. The lines with ``circles" and
``triangles" in Fig. \ref{Fig:order-parameter}(b) represent the
dependence of $R_{1}$ and $R_2$ on $P$, respectively. It is easy
to see that both $R_{1}$ and $R_2$ increase with $R$ and are much
larger than the corresponding total $R$. Especially, $R_{1}$
will be close to unity when $P>0.6$, indicating that most of the
conformists has become synchronized. To understand the relationship
between the conformists and contrarians, Fig. \ref{Fig:phase}
shows how the parameter $P$ influences the evolution of the
average phases $\Psi_1$ and $\Psi_2$ in Eqs. (\ref{order-conformist})
and (\ref{order-contrarian}) and where (a)-(c) represent the
cases of $P=0, 0.5$ and $1$, respectively, and (d) represents the
dependence of $\Delta \Psi$ on $P$ with $\Delta \Psi=\langle
|\Psi_1-\Psi_2|\rangle$. From Fig. \ref{Fig:phase}(d) we see that
$\Delta \Psi$ increases with $P$ and approaches $\pi$ when $P$ is
relatively large, indicating that the tit-for-tat strategy has
made the conformists and contrarians become two independent
synchronized groups with a phase difference close to $\pi$. Substituting
this result into Eq. (\ref{order}), we find that the two groups
of conformists and contrarians will have the opposite contributions
to the order parameter $R$ and thus make the value of $R$ in
Fig. \ref{Fig:order-parameter}(b) much less than the corresponding
$R_1$ and $R_2$.

\begin{figure}
\epsfig{figure=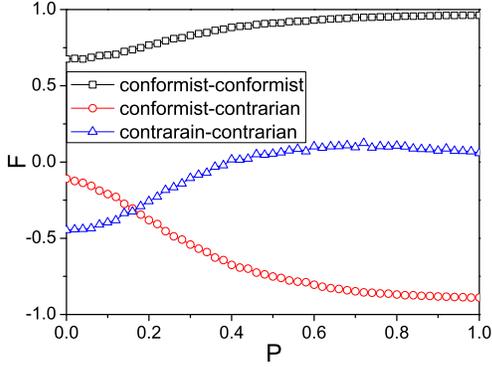,width=1.0\linewidth} \caption{(color
online). Frustration $F$ versus $P$ where the three curves with
``squares", ``circles" and ``triangles" represent the frustration
on the links of conformist-conformist, conformist-contrarian, and
contrarian-contrarian, respectively. } \label{Fig:frustration}
\end{figure}

To understand the phase difference $\Delta \Psi$ better, we introduce
the link frustration $f_{ij}$ for a connected link $i-j$ as follows
\cite{Zanette:2005,Levnajic:2011}
\begin{equation}
\label{frust} f_{ij}=\cos(\theta_i-\theta_j),
\end{equation}
which represents the local dynamical information. Depending on the
property of nodes $i$ and $j$, $f_{ij}$ can be classified into three
classes of conformist-conformist, conformist-contrarian, and contrarian-contrarian.
Those links of conformist-contrarian denote the connections between the two
groups of conformists and contrarians and thus represent the asymmetric
interaction between the positive and negative couplings. The global
frustration can be defined as the network average of $f_{ij}$
\begin{equation}
\label{frust1} F=\frac{\sum_{i=1}^N\sum_{j=1}^N
|A_{ij}|f_{ij}}{\sum_{i=1}^N\sum_{j=1}^N |A_{ij}|}.
\end{equation}
Fig.\ref{Fig:frustration} shows the dependence of $F$ on $P$ where
the three curves with ``squares", ``circles" and ``triangles"
represent the frustration on the links of conformist-conformist,
conformist-contrarian, and contrarian-contrarian, respectively.
From Fig.\ref{Fig:frustration} we see that with the increase of
$P$, $F$ will approach $1,-1,$ and $0$ for the three classes of
conformist-conformist, conformist-contrarian, and
contrarian-contrarian, respectively. It is not difficult to
understand the two cases of $F\approx 1$ and $-1$ as we have
$\Delta \Psi\approx 0$ for the links of conformist-conformist with
attractive coupling and $\Delta \Psi\approx \pm \pi$ for the links
of conformist-contrarian with repulsive coupling. The
case of $F\approx 0$ implies $\Delta \Psi\approx \pm \pi/2$, indicating
that the group of contrarian-contrarian is not highly synchronized.

\begin{figure}
\epsfig{figure=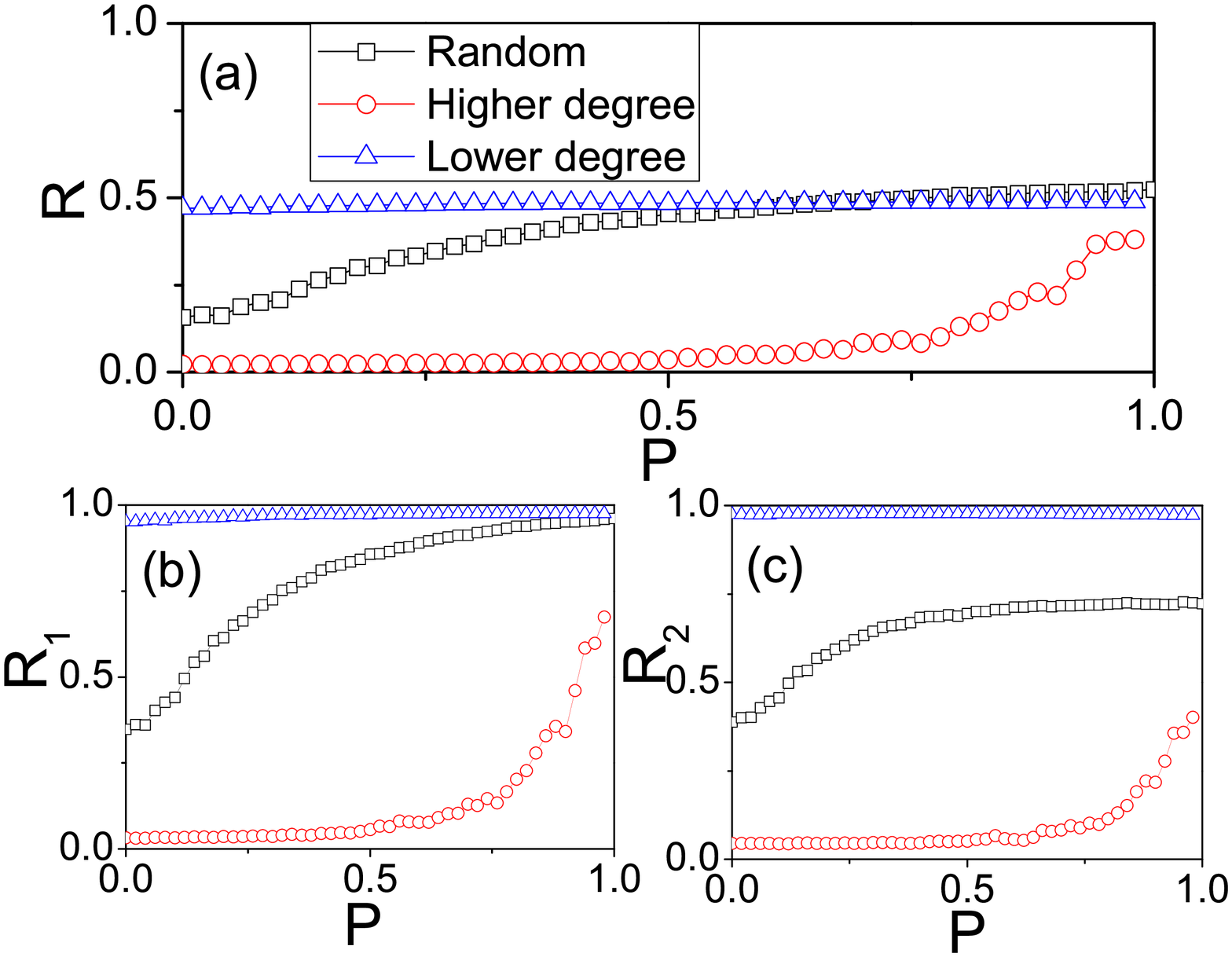,width=1.0\linewidth} \caption{(color
online). Case of BA network corresponding to Fig.
\ref{Fig:order-parameter}(b) where the three curves with
``squares", ``circles" and ``triangles" represent the cases of
choosing nodes of contrarians by random, the higher degrees and
lower degrees, respectively. (a) $R$ versus $P$; (b) $R_1$ versus
$P$; (c) $R_2$ versus $P$. } \label{Fig:BA}
\end{figure}

To check the influence of network topology, we now turn to a
scale-free network according to the algorithm given by
Barabasi-Albert (BA) \cite{Barabasi:1999}, where the other
parameters are remained the same as in the case of ER network. In
contrast to the homogeneous degree distribution in the ER network,
the BA network is heterogeneous with a power-law degree
distribution. Thus, except the random uniformly choosing the
contrarian nodes, we also consider the cases of choosing only
those nodes with larger degrees or smaller degrees as contrarians.
Fig. \ref{Fig:BA} shows the results where the three curves with
``squares", ``circles" and ``triangles" represent the cases of
choosing nodes of contrarians by random, the higher degrees and
lower degrees, respectively, and (a)-(c) denote the cases of $R$,
$R_1$ and $R_2$, respectively. Comparing the curves with
``squares" in Fig. \ref{Fig:BA}(a)-(c) with the corresponding
curves in Fig. \ref{Fig:order-parameter}(b), respectively, we find
that they are similar, indicating that the tit-for-tat strategy
also works for the BA network. Then we notice that all the curves
with ``triangles" in Fig. \ref{Fig:BA}(a)-(c) are approximate
constant, implying that the nodes with lower degrees are not very
important. However, we also notice that all the curves with
``circles" in Fig. \ref{Fig:BA}(a)-(c) are sensitive on $P$ only
when $P>0.5$. The mechanism can be explained as follows. A
characteristic feature of the BA network is that the hubs have a
large number of links while the leaves have a few links. When the
contrarians are chosen from the leaves, the links of conformist-contrarian
will be small and thus their influence through the tit-for-tat strategy
is also small, resulting the approximate constant curves with
``triangles" in Fig. \ref{Fig:BA}. When the contrarians are chosen
from the hubs, the links of conformist-contrarian will be large.
However, as most of the neighbors of a hub are leaves, it is very
possible for those links of hub-leaves to be chosen for the tit-for-tat
strategy when $P<0.5$. We know that the leaves will not take an important
role in the collective behaviors, thus resulting in the insensitive on
$P$ in the curves with ``circles" in Fig. \ref{Fig:BA} for $P<0.5$.
When $P>0.5$, the possibility to choose those links of hub-hub or
hub-middle nodes for the tit-for-tat strategy will gradually increase,
resulting in the sensitive on $P$ in the curves with ``circles" in
Fig. \ref{Fig:BA} for $P>0.5$. Similarly, we can understand the case of
choosing the contrarians randomly, i.e. the curves with ``squares" in
Fig. \ref{Fig:BA}.

In sum, the case of random uniformly distributed
contrarians is much more sensitive to $P$, in contrast to the two
cases of choosing the nodes of higher and lower degrees as
contrarians, indicating that the tit-for-tat strategy may be more
efficient when the contrarians are random uniformly mixed. This
result is a little counterintuitive as we usually think that it
would be more efficient to choose the nodes of higher degrees as
contrarians. Thus, this finding is of interesting as it may have
some applications. For example, in the case of epidemic spreading,
people are more easy to be vaccinated when the contrarians are
random distributed. For the same reason, in the political
election, a final winner can be easier to be generated when the
contrarians are random distributed.

\section{Case II: contrarian oscillators sending negative coupling to
their neighbors}

In this case, Eq. (\ref{Kuramoto}) will be replaced by the following equation
\begin{equation}
\label{Kuramoto1}
\dot{\theta_{i}}=\omega_i+\lambda\sum_{j=1}^NB_{ij}\sin(\theta_j-\theta_i),
\quad i=1,\ldots,N
\end{equation}
where $\lambda$ is the coupling strength of all the oscillators,
and $B_{ij}=1$ when the connected neighboring node $j$ is conformist
and $B_{ij}=-1$ when the connected neighboring node $j$ is
contrarian. A distinguished difference between Eq.
(\ref{Kuramoto}) and Eq. (\ref{Kuramoto1}) is that in Eq.
(\ref{Kuramoto}), all the neighbors of node $i$ are in the same position.
While in Eq. (\ref{Kuramoto1}), the neighbors of node $i$ are in different
positions, i.e. the conformist neighbors have positive contribution while the
contrarian neighbors have negative contribution. We keep the other
parameters and the tit-for-tat strategy the same as in {\em Case
I}. That is, we let all the asymmetric couplings of
$B_{ij}=-B_{ij}$ become symmetric ones of $B_{ij}=B_{ij}=-1$ by a
probability $P$, i.e. the tit-for-tat strategy.

\begin{figure}
\epsfig{figure=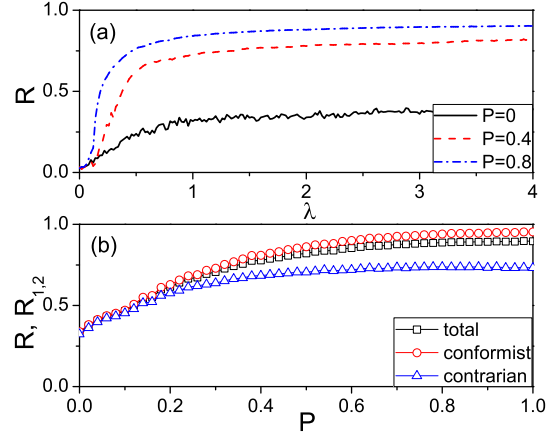,width=1.0\linewidth} \caption{(color
online). {\em Case II}: Synchronization diagrams by the
tit-for-tat strategy. (a) $R$ versus $\lambda$ where the ``solid",
``dashed", and ``dash-dotted" lines represent the cases of $P=0,
0.4$, and $0.8$, respectively. (b) $R$ and $R_{1,2}$ versus $P$
for $\lambda=3.0$, where the lines with ``squares", ``circles",
and ``triangles" represent the cases of the total oscillators, the
group of conformists, and the group of contrarians, respectively.}
\label{Fig:order-parameter1}
\end{figure}

In numerical simulations, we first take the random
ER network with the same parameters as in {\em Case I} as an
example. Fig. \ref{Fig:order-parameter1} shows the results
corresponding to Fig. \ref{Fig:order-parameter}. From Fig.
\ref{Fig:order-parameter1}(b) we see that all the three curves
gradually increase with $P$, indicating the efficiency of the
tit-for-tat strategy. Comparing Fig. \ref{Fig:order-parameter}
with Fig. \ref{Fig:order-parameter1}, we see that they have two
differences: (1) All the curves in Fig. \ref{Fig:order-parameter1}
are higher than the corresponding ones in Fig.
\ref{Fig:order-parameter}, indicating that the tit-for-tat
strategy is more effective in {\em Case II} than in {\em Case I}.
(2) When $P<0.2$, all the three curves in Fig.
\ref{Fig:order-parameter1} are overlapped but separated in Fig.
\ref{Fig:order-parameter}. This is because the coupling term in
Eq. (\ref{Kuramoto1}) is only determined by the neighbors of node
$i$ no matter node $i$ is a conformist or a contrarian. When
$P=0$, Eq. (\ref{Kuramoto1}) will be statistically the same for
both a conformist and a contrarian and thus results in the overlap
in Fig. \ref{Fig:order-parameter1}(b). While the coupling term in
Eq. (\ref{Kuramoto}) is determined not only by the neighbors of
node $i$ but also by node $i$ itself, which results in a
difference between the conformists and the contrarians and thus
the separation in Fig. \ref{Fig:order-parameter}(b).

\begin{figure}
\epsfig{figure=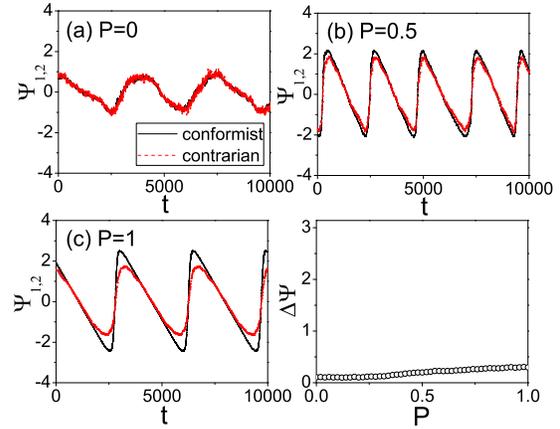,width=1.0\linewidth} \caption{(color
online). {\em Case II}: Evolution of the average phases $\Psi_1$
and $\Psi_2$ where (a)-(c) represent the cases of $P=0, 0.5$ and
$1$, respectively, and (d) represents the dependence of $\Delta
\Psi$ on $P$.} \label{Fig:phase1}
\end{figure}

Fig. \ref{Fig:phase1} shows the influence of $P$ on the evolution
of the average phases $\Psi_1$ and $\Psi_2$, corresponding to Fig.
\ref{Fig:phase}. Comparing the corresponding curves in Fig.
\ref{Fig:phase1} with that in Fig. \ref{Fig:phase}, we observe
that $\Delta \Psi$ is close to zero in Fig. \ref{Fig:phase1} but
close to $\pi$ in Fig. \ref{Fig:phase}, indicating that the
conformists and contrarians will become two independent
synchronized groups in {\em Case I} but stay in the same
synchronized group in {\em Case II}.

\begin{figure}
\epsfig{figure=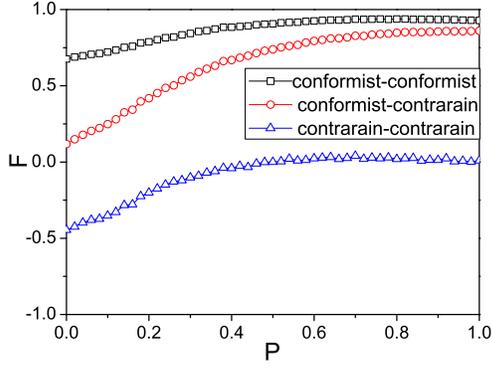,width=1.0\linewidth} \caption{(color
online). {\em Case II}: Frustration $F$ versus $P$ where the three
curves with ``squares", ``circles" and ``triangles" represent the
frustration on the links of conformist-conformist,
conformist-contrarian, and contrarian-contrarian, respectively. }
\label{Fig:frustration1}
\end{figure}

Fig.\ref{Fig:frustration1} shows the dependence of $F$ on $P$,
corresponding to Fig.\ref{Fig:frustration}. From
Fig.\ref{Fig:frustration1} we see that with the increase of $P$,
$F$ will approach $1,1,$ and $0$ for the three classes of
conformist-conformist, conformist-contrarian, and
contrarian-contrarian, respectively. Comparing
Fig.\ref{Fig:frustration1} with Fig.\ref{Fig:frustration} we see
that the two classes of conformist-conformist and
contrarian-contrarian are similar between the {\em Case I} and
{\em Case II} while the class of conformist-contrarian changes
from $F\approx -1$ in Fig.\ref{Fig:frustration} to $F\approx 1$ in
Fig.\ref{Fig:frustration1}. This point can be understood as
follows: In {\em Case I}, the formed two synchronized groups have
a phase difference $\pi$ and thus the two nodes of a link with
conformist-contrarian have a phase difference $\pi$ or opposite
phase, which results in $F\approx -1$. While in {\em Case II}, both
the conformists and contrarians are in the same synchronized group,
which results in $F\approx 1$.

\begin{figure}
\epsfig{figure=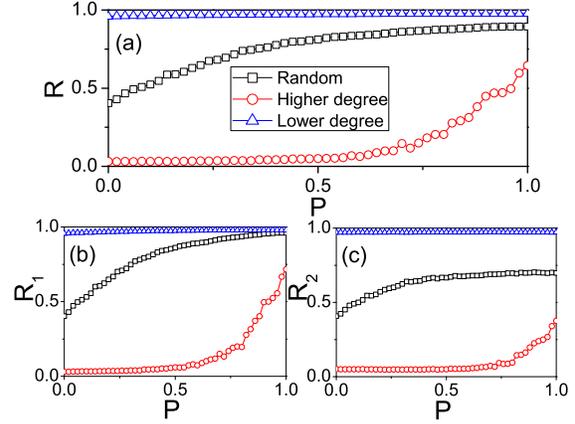,width=1.0\linewidth} \caption{(color
online). {\em Case II}: Case of BA network corresponding to Fig.
\ref{Fig:order-parameter1}(b) where the three curves with
``squares", ``circles" and ``triangles" represent the cases of
choosing nodes of contrarians by random, the higher degrees and
lower degrees, respectively. (a) $R$ versus $P$; (b) $R_1$ versus
$P$; (c) $R_2$ versus $P$. } \label{Fig:BA1}
\end{figure}

Fig. \ref{Fig:BA1} shows the results of BA network corresponding
to Fig. \ref{Fig:BA}, where the three curves with ``squares",
``circles" and ``triangles" represent the cases of choosing nodes of
contrarians by random, the higher degrees and lower degrees,
respectively. Comparing the three panels of Fig. \ref{Fig:BA1} with
the corresponding ones of Fig. \ref{Fig:BA}, we see that the
corresponding panels (b) and (c) are similar each other while all the
curves in panel (a) of Fig. \ref{Fig:BA1} are approximate two times
of that in panel (a) of Fig. \ref{Fig:BA}, confirming that (1) there
is one synchronized group in {\em Case II} and two synchronized
groups in {\em Case I}; and (2) the tit-for-tat strategy is more
efficient for the case of random uniformly distributed contrarians.

\section{A brief theoretical analysis}
To understand the role of tit-for-tat strategy better, we here present a brief
theoretical analysis. For the convenience of analysis, we rewrite
Eq. (\ref{Kuramoto}) and Eq. (\ref{Kuramoto1}) in a unified framework
\begin{equation}
\label{Kuramoto2}
\dot{\theta_{i}}=\omega_i+\lambda\sum_{j=1}^NC_{ij}\sin(\theta_j-\theta_i),
\quad i=1,\ldots,N
\end{equation}
with $\lambda>0$. For {\em Case I}, we have $C_{ij}=A_{ij}=1$ when node
$i$ is a conformist and $C_{ij}=-A_{ij}=-1$ when node $i$ is a contrarian.
For {\em Case II}, we always have $C_{ij}=B_{ij}$. We follow the
Ref. \cite{Restrepo:2005} to introduce a local order parameter $r_i$ to
quantify the coherence of the inputs to a given node $i$, which is defined by
\begin{equation}
\label{order2} r_ie^{i\Psi_i}=\sum_{j=1}^N C_{ij}\langle e^{i\theta_j}\rangle_t,
\end{equation}
where $\langle \cdots\rangle_t$ denotes a time average. The right side of
Eq. (\ref{order2}) can be divided into two parts: one from the conformist
neighbors and the other from the contrarian neighbors. Thus, we rewrite
Eq. (\ref{order2}) as
\begin{equation}
\label{order3} r_ie^{i\Psi_i}=\sum_{j\in conformists} C_{ij}\langle e^{i\theta_j}\rangle_t
+\sum_{j\in contrarians} C_{ij}\langle e^{i\theta_j}\rangle_t
\end{equation}

\begin{figure}
\epsfig{figure=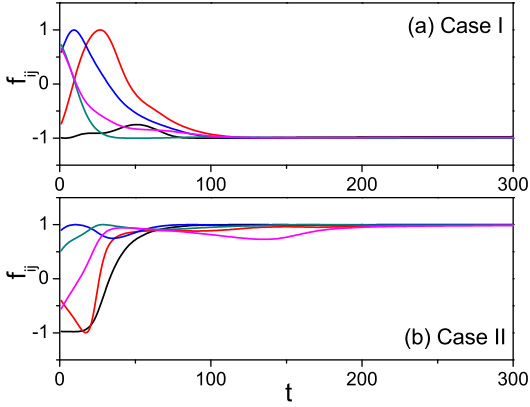,width=1.0\linewidth} \caption{(color
online). Evolution of the link frustration $f_{ij}$ for several typical
links of conformist-contrarian in the ER network when $P=1$.
(a) $f_{ij}$ versus $t$ for {\em Case I}.
(b) $f_{ij}$ versus $t$ for {\em Case II}. } \label{Fig:link}
\end{figure}

The links of conformist-contrarian are the key elements in Eq.
(\ref{order3}) as the tit-for-tat strategy is applied only on
them. We here take the case of $P=1$ as an example. For {\em Case
I}, we first discuss the case of a conformist node $i$. Before
using the tit-for-tat strategy, all the $C_{ij}$ in Eq.
(\ref{order3}) are positive. After using the tit-for-tat strategy,
the $C_{ij}$ in the second term of the right side of Eq.
(\ref{order3}) will become negative. The competition between the
two terms in the right side of Eq. (\ref{order3}) will depend on
their phase differences, which can be reflected by the link
frustration of Eq. (\ref{frust}). To figure out the phase
differences, Fig. \ref{Fig:link}(a) shows the evolution of the
link frustration $f_{ij}$ with $P=1$ for several typical links of
conformist-contrarian in the ER network. From Fig.
\ref{Fig:link}(a) we see that after the transient process, all the
$f_{ij}$ will become approximately $-1$, implying that the two
terms in the right side of Eq. (\ref{order3}) will have a phase
difference $\pi$. That is, the two terms in the right side of Eq.
(\ref{order3}) become two synchronized groups with a phase
difference $\pi$. Considering one more fact that the values of
$C_{ij}$ in the two synchronized groups are $\pm 1$, we see that
the negative $C_{ij}$ and the phase difference $\pi$ in the second
term of the right side of Eq. (\ref{order3}) will make it have the
same positive contribution as the first term of the right side of
Eq. (\ref{order3}), and thus enhance the local order parameter
$r_i$.

The enhanced $r_i$ will then make the $r_i$ of a contrarian node
$i$ increase. That is, all the $r_i$ will increase with $P$. In
the global level, the order parameter $R$ can be expressed as
\cite{Restrepo:2005}
\begin{equation}
\label{order1}
R=\frac{\sum_{i=1}^N r_i}{\sum_{i=1}^N\sum_{j=1}^N |A_{ij}|}
\end{equation}
Thus, the enhanced $r_i$ will make $R$ increase with $P$, which explains the
numerical results of Fig. \ref{Fig:order-parameter}.

For {\em Case II}, we first discuss the case of a contrarian node
$i$. Before using the tit-for-tat strategy, the $C_{ij}$ in the
first term of the right side of Eq. (\ref{order3}) is positive
while the $C_{ij}$ in the second term of the right side of Eq.
(\ref{order3}) is negative. After using the tit-for-tat strategy,
the $C_{ij}$ in the first term of the right side of Eq.
(\ref{order3}) will become negative, i.e. all the $C_{ij}$ in Eq.
(\ref{order3}) are negative now. To figure out the phase
differences between the two terms of the right side of Eq.
(\ref{order3}), Fig. \ref{Fig:link}(b) shows the evolution of the
link frustration $f_{ij}$ with $P=1$ for several typical links of
conformist-contrarian in the ER network. From Fig.
\ref{Fig:link}(b) we see that after the transient process, all the
$f_{ij}$ will become approximately $+1$, implying that the two
terms in the right side of Eq. (\ref{order3}) will have no phase
difference. That is, the two terms in the right side of Eq.
(\ref{order3}) become one synchronized group and thus have the
same contribution to the synchronization, which enhances the
local order parameter $r_i$. The enhanced $r_i$ will then make the
$r_i$ of a conformist node $i$ increase. That is, all the $r_i$
will increase with $P$ and thus make $R$ of Eq. (\ref{order1})
increase with $P$, which explains the numerical results of Fig.
\ref{Fig:order-parameter1}.

\section{Discussions and conclusions}

\begin{figure}
\epsfig{figure=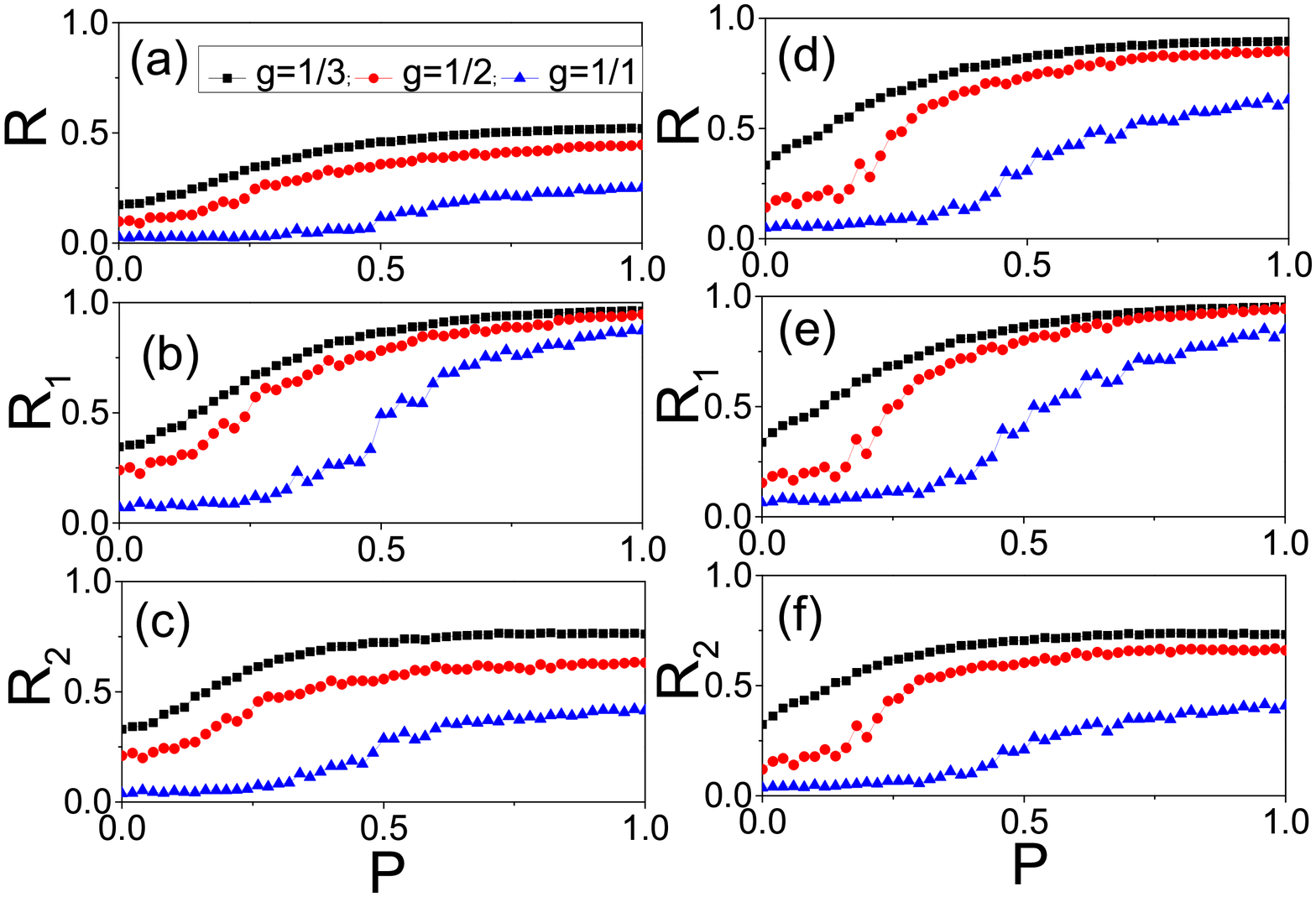,width=1.0\linewidth} \caption{(color
online). How the ratio $g$ between the contrarians and conformists influences
$R, R_1$ and $R_2$ where the ``squares", ``circles" and ``triangles"
represent the cases of $g=1/3, 1/2$ and $1/1$, respectively,
and (a)-(c) denote the {\em Case I} and (d)-(e) denote the
{\em Case II}. } \label{Fig:ratio}
\end{figure}

Although all the above discussions are based on $\rho=0.25$, we find that
the obtained results also work for other $\rho$. To show this point in detail,
we let $g$ be the ratio between the contrarians and conformists, which gives
$g\equiv\frac{\rho N}{(1-\rho)N}=\frac{1}{3}$ for $\rho=0.25$. Figure \ref{Fig:ratio}
shows how $g$ influences $R, R_1$ and $R_2$ where the ``squares", ``circles"
and ``triangles" represent the cases of $g=1/3, 1/2$ and $1/1$, respectively,
and (a)-(c) denote the {\em Case I} and (d)-(e) denote the {\em Case II}. From
Fig. \ref{Fig:ratio} we see that all the three quantities $R, R_1$ and $R_2$
increase with $P$, no matter it is the {\em Case I} or {\em Case II}. Thus,
the tit-for-tat strategy also works for other $g$ or $\rho$. On the other hand,
from Fig. \ref{Fig:ratio} we notice that all the curves with $g=1/3$ are above
than that of $g=1/2$ and then both of them are above that of $g=1/1$, indicating
that $R, R_1$ and $R_2$ decrease with the increase of $g$.

The strategy of tit-for-tat is widely used in animal and social
competition and is proved to be an efficient way to reach a result
of cooperation. The explanation of its microscopic mechanism
results in a famous minority game theory based on payoff. In this
paper, we first time apply this strategy to the networked systems
of coupled oscillators. We reveal that the tit-for-tat strategy
takes effect through two elements of the links of
conformist-contrarian, i.e. the phase difference and symmetry of
the two couplings. The degree of the variation of these two
elements will result in different order parameter $R$ for
synchronization. This finding is of meaningful in explaining how
the neurons network with excitable and inhibitory neurons
implements a diversity of cognitive processes, which is closely
related to the degree of synchronization.

We notice from both Fig.\ref{Fig:frustration} and
Fig.\ref{Fig:frustration1} that the value of $F$ for the links of
contrarian-contrarian is approximately zero when $P>0.5$,
indicating that the phase difference between two contrarians is
about $\pi/2$. This result tells us that a contrarian is always
repulsive and thus it is difficult to reach a complete
synchronized group of contrarians with $R_2=1$. Therefore, an
effective way to increase $R$ is to reduce the fraction of
contrarian-contrarian interaction, i.e. avoiding two contrarians
to be the nearest neighbors.

In conclusions, we have studied the synchronization of a networked
system with both positive and negative couplings. We reveal that
the negative role of repulsive oscillators to synchronization can
be effectively reduced by applying a tit-for-tat strategy. Two
kinds of negative couplings are considered and it is found that
the tit-for-tat strategy works for both of them. The underlying
mechanism is to use the tit-for-tat strategy to against the phase
difference of the links of conformist-contrarian and thus increase
the synchronization. This result may provide new insights to the
diversity of cognitive processes.

This work was partially supported by the NNSF of China under Grant
No. 11135001 and 973 Program under Grant No. 2013CB834100.

\end{document}